\documentclass{jfm}
\usepackage{graphicx}
\usepackage{epstopdf, epsfig}
\usepackage{gensymb}
\usepackage{subcaption}
\usepackage[dvipsnames]{xcolor}

\shorttitle{LCSs and entrainment in gravity currents}
\shortauthor{Marius M. Neamtu-Halic, Dominik Krug, George Haller and Markus Holzner}

\title{Lagrangian coherent structures and entrainment near the turbulent/non-turbulent interface of a gravity current}

\author{Marius M. Neamtu-Halic\aff{1}
  \corresp{\email{neamtu@ifu.baug.ethz.ch}},
  Dominik Krug\aff{2},
  George Haller\aff{3}
 \and  Markus Holzner\aff{1}}

\affiliation{\aff{1}Institute of Environmental Engineering, ETH Zürich, CH-8039 Zürich, Switzerland
\aff{2}Physics of Fluids Group and Twente Max Planck Center, Department of Science and Technology, Mesa+ Institute, and J.M. Burgers Center for Fluid Dynamics, University of Twente, PO Box 217, 7500 AE Enschede, The Netherlands
\aff{3}Institute of Mechanical Systems, ETH Zürich, 8092 Zürich, Switzerland}

\begin{document}

\maketitle

\begin{abstract}

In this paper, we employ Lagrangian coherent structures (LCSs) theory for the three-dimensional vortex eduction and investigate the effect of large-scale vortical structures on the turbulent/non-turbulent interface (TNTI) and entrainment of a gravity current. The gravity current is realized experimentally and different levels of stratification are examined. For flow measurements, we use a multivolume three-dimensional particle tracking velocimetry technique. To identify vortical LCSs (VLCSs), a fully automated 3D extraction algorithm for multiple flow structures based on the so-called Lagrangian-Averaged Vorticity Deviation method is implemented. The size, the orientation and the shape of the VLCSs are analyzed and the results show that these characteristics depend only weakly on the strength of the stratification. Through conditional analysis, we provide evidence that VLCSs modulate the average TNTI height, affecting consequently the entrainment process. Furthermore, VLCSs influence the local entrainment velocity and organize the flow field on both the turbulent and non-turbulent sides of the gravity current boundary. 

\end{abstract}

\begin{keywords}
Stratified turbulence, gravity currents, Lagrangian coherent structures 
\end{keywords}
\section{Introduction}
The flow in the vicinity of the sharp interface that is widely observed to form between a turbulent flow and non-turbulent surroundings, e.g. a chimney plume issuing into quiescent air, has received considerable attention in the literature over the last decades \citep[e.g.][]{dimotakis2000mixing, holzner2008lagrangian, da2014interfacial}. Among others, the main motivation for these studies stems from the fact that across turbulent/non-turbulent interfaces (TNTIs), fluid is continuously incorporated into the turbulent flow, a process known as turbulent entrainment. The entrainment rate has direct bearing on mixing properties and global dynamics of the flow and is therefore of high relevance and interest in many  applications, e.g. jet-, wake- and boundary layer-flows.\\ 
To date, much research has focused on small-scale properties of the entrainment process \citep[see e.g.][]{ westerweel2005mechanics, holzner2011laminar, silva2018scaling} and it is now well established \citep{mathew2002some, westerweel2009momentum}, that the process by which non-turbulent fluid initially acquires vorticity is of viscous nature, as originally envisioned by \citet{corrsin1955free}. However, the overall entrainment rate is known to be independent of viscosity or, in other terms, of the Reynolds number \citep[see e.g.][]{tritton1988physical, tsinober2009informal}. It is therefore believed that structures at larger scales act to cancel the Reynolds number dependence of the small-scale process. That is to say that, even though locally non-turbulent fluid becomes turbulent via viscous diffusion of vorticity, the overall entrainment is imposed by fluid motion at larger scales which control the surface area of the TNTI \citep{townsend1980structure, sreenivasan1989mixing}. Recently, \citet{lee2017signature} used conditional analysis to show that the TNTI surface area of a turbulent boundary layer increases in the vicinity of large-scale motions (LSMs). However, a similar observation is missing for other flows, far from the wall, and nowadays it is not clear how the fluid motion near the TNTI is related to the vortical structures in the flow.\\ 
\citet{da2011role} visualized the vortical structures near the TNTI of a turbulent planar jet. They suggested that the large-scale vortices “sitting” on the TNTI are mostly defining its shape. Moreover, they conclude that the characteristic vorticity jump of the TNTI, as well as its thickness, is imposed by the radial vorticity distribution of these structures. \\
Nevertheless, progress in our understanding of the relation between the large-scale vortical structures and TNTI has been hampered by the arbitrariness in the ‘vortex’ structure definition. Often structures are extracted based on arbitrary thresholds and based on quantities that are not invariant to a change of system of reference, i.e. they are not objective. Newly developed, Lagrangian methods \citep[for a review, see][]{haller2015lagrangian} for vortex identification constitute a promising tool to overcome this issue. \\
Since the initial work of \citet{haller2000lagrangian}, the Lagrangian coherent structures (LCSs) theory aims to identify vortical structures (referred hereinafter as VLCSs to distinguish them by other type of LCSs) using dynamical systems approaches, overcoming the arbitrariness that characterizes the classical non-objective methods, such as $Q-$ \citep{hunt1988eddies}, ${\Delta}-$ \citep{chong1990general} and ${\lambda}_{2}-criterion$ \citep{jeong1995identification}. LCSs approaches are mostly based on stretching requirements \citep{haller2015lagrangian} and identify highly coherent, ‘black-hole’ type material regions with high accuracy, but at substantial computational cost \citep[see e.g.][]{haller2013coherent,hadjighasem2016geodesic}.\\ 
Recently, a less computationally expensive approach has been developed that replaces the stretching-based coherence requirement with rotational coherence. This method uses a new dynamic version of the classic polar decomposition introduced in \citet{haller2016dynamic} and identifies the initial positions of VLCSs as tubular level surfaces of the so-called Lagrangian-Averaged Vorticity Deviation (LAVD). \citet{haller2016defining} identified vortical structures, using LAVD-based methods, in two-dimensional and three-dimensional flow fields. However, as highlighted by \citet{haller2016defining}, a fully automated implementation of LAVD methods for multiple three-dimensional coherent structures is still missing.\\
In the present work, we seek to implement a three-dimensional VLCS extraction method based on the LAVD theory of \citet{haller2016defining} and apply it to experimental data of a gravity current. The gravity current constitutes an interesting flow case for two reasons. On one hand, it has important practical applications, e.g. river plumes, katabatic winds and oceanic overflows. On the other hand, the entrainment rate across the TNTI varies with the ratio between the buoyancy and the flow shear strengths, represented by the Richardson number, $Ri$. This allows us to investigate how the properties of the TNTI vary to accommodate the entrainment variation with $Ri$ and how these properties are related to the VLCSs in the proximity of the TNTI.\\ 
The paper is organized as follows. Section \ref{sec:methods} describes the experimental measurements, together with the TNTI identification and VLCSs eduction methods. Section \ref{sec:results} characterizes the VLCSs and analyzes their relationship with the TNTI and the entrainment process. The article closes with the discussion and conclusions in section \ref{sec:discussion}.
\section{Methods}\label{sec:methods}
\subsection{Experiments}
\begin{figure}
  \centerline{\includegraphics[width=1\linewidth]{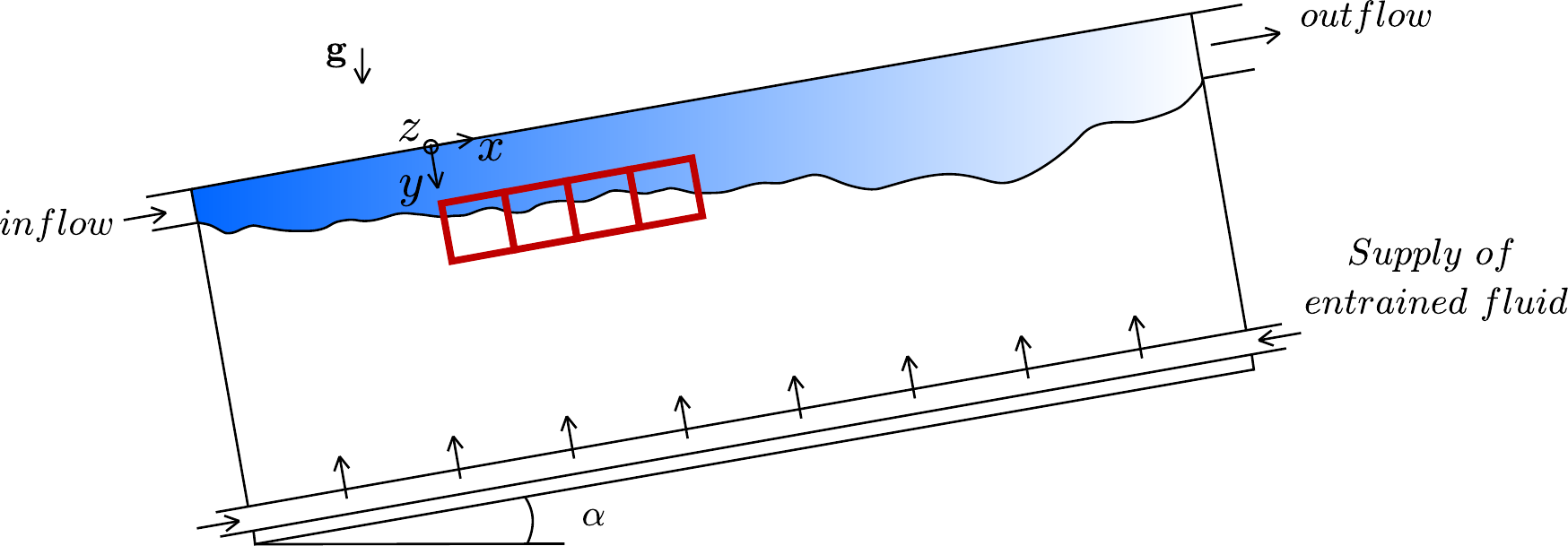}}
  \caption{Sketch of the experimental setup. The blue area indicates the gravity current (lighter turbulent fluid) that flows along the top wall of the tank. At the bottom, heavy fluid is resupplied  to make up for the entrained flux, while mixed fluid spills out of the tank through the outlet at the top-end of the tank. The four 3D-PTV investigation domains are indicated by red rectangles.}
\label{fig:fig1}
\end{figure}

The gravity current data presented here were collected using the experimental apparatus developed in \citet{krug2013experimental}. This setup is sketched in figure \ref{fig:fig1} and was designed to create a gravity current along the top of an inclined glass tank, which can be tilted between 0 and 90 degrees and whose dimensions are 2 $m$ long and 0.5 $m$ wide and high.  The gravity current was realized by the continuous injection of a light fluid (a mixture of water and ethanol) along the top wall of the tank into a denser ambient fluid (a mixture of water and sodium chloride). As outlined in detail in \citet{krug2014combined}, a proper preparation of the solutions provides the desired density difference, while keeping the same refractive index in the two solutions. The latter is a crucial requirement for optical measurement techniques. During the experiment, the flow rate of the light fluid is driven by a water pump, measured via a flowmeter and its feedback is implemented as a closed loop control. In this way, a constant flow rate is guaranteed throughout the entire experiment. The natural transition to turbulence of the light fluid via Kelvin-Helmholtz instabilities requires an impracticably long tank \citep[for a discussion, see][]{krug2013experimental}. It was therefore preferred to force the transition to turbulence at the inlet by means of a diffuser equipped with rotating flapping grids. In previous experimental studies by \citet{krug2013experimental} and \citet{odier2014entrainment}, it was shown that with this system the turbulence characteristics at a location sufficiently ''far'' from the inlet, as in the case of the present study, are independent on the inflow turbulence. The ambient entrained fluid was gently resupplied along the bottom of the tank to replenish entrained fluid. As noted by \citet{krug2013experimental} the particular value of the flow rate of the ambient fluid does not influence the entrainment rate, however a proper choice of it permits to avoid large-scale recirculation within the tank.\\
In this paper, we present results for three different flow cases. They differ in the initial density difference between the two solutions $\Delta\rho_{0}$ and the tank inclination $\alpha$. An overview of the flow parameters is presented in Table \ref{table:tab1}. To compute the inflow Reynolds number, $Re_{0}$, and the inflow Richardson number, $Ri_{0}$, we used the inlet height $d$ and the mean inflow velocity $U_{0}$. Note that the label of the flow cases designates the respective value of $Ri_{0}$. As shown by \citet{ellison1959turbulent}, a gravity current adjusts itself to an equilibrium $Ri$ number that depends only on the inclination of the tank $\alpha$. Recently, \citet{negretti2017development} demonstrated that for a gravity current at the onset of the turbulence, the equilibrium Richardson number depends also on the inflow interfacial Richardson number. Maintaining a constant inflow velocity $U_{0}$, we varied $\Delta\rho_{0}$ such that the flow results to be close to the equilibrium state near the inflow. This was guided by the numerical results of \citet{krug2017fractal} and \citet{van2018mixing}.
\subsection{Measurements}
\begin{figure}
  \centerline{\includegraphics[width=1\linewidth]{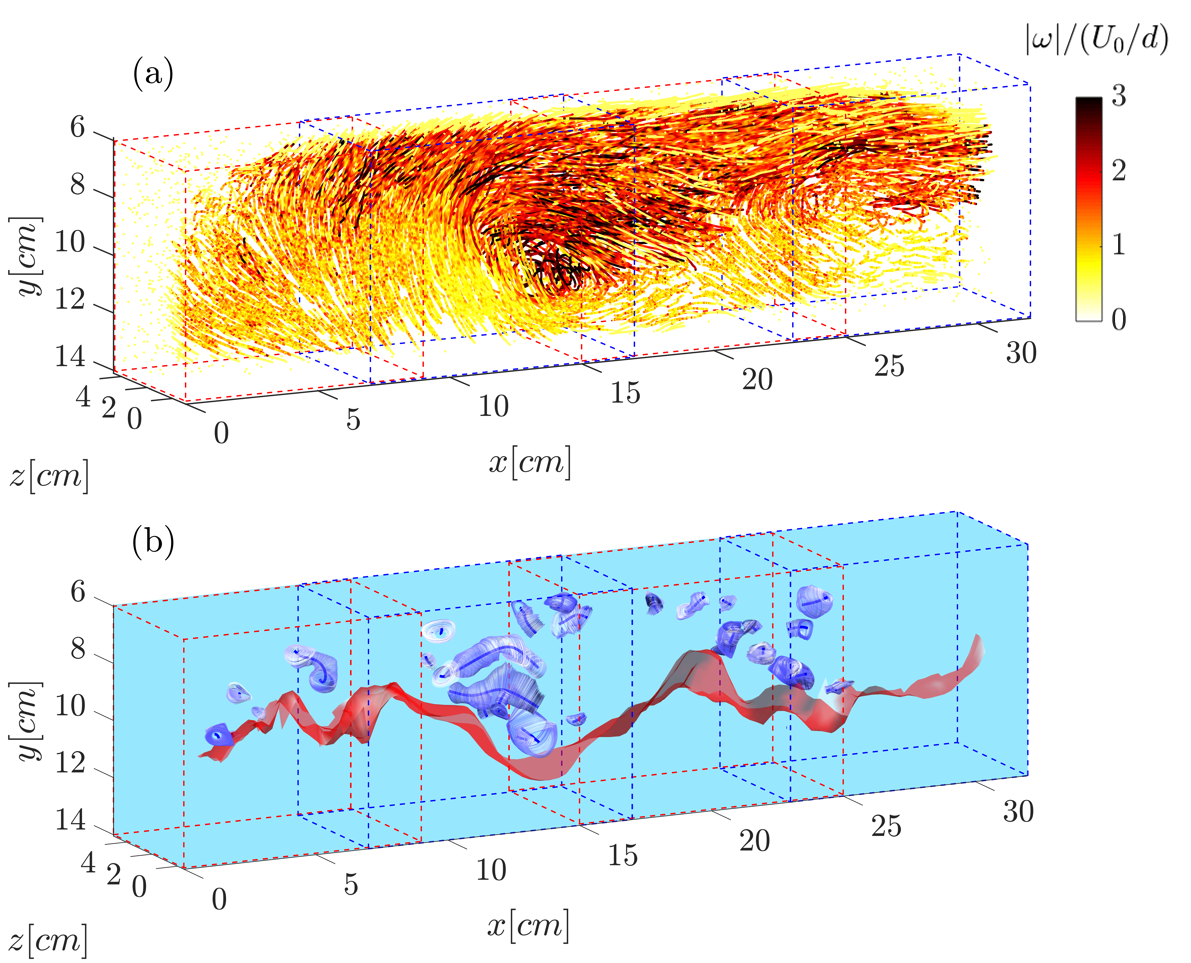}}
  \caption{(a) Three-dimensional fluid trajectories color-coded with the norm of the vorticity for the flow case Ri20. The time period shown here is equivalent to three turnover-times (defined in section \ref{subsection:extraction}) of the largest eddies. The four alternated red and blue rectangular outlines represent the four 3D-PTV observation volumes. (b) Corresponding three-dimensional VLCSs, represented by blue tubular surfaces (boundaries) surrounding one-dimensional curves (centers), and the TNTI of the gravity current (red open-surface). The region above the TNTI is turbulent, whereas below the flow is irrotational.}
\label{fig:fig2}
\end{figure}

Flow measurements were performed using three-dimensional particle tracking velocimetry (3D-PTV). In order to capture a large investigation volume while maintaining a fine spatial resolution, which is crucial for the VLCSs extraction method used here, we performed measurements using four separate 3D-PTV systems. Their individual measurement domains were then stitched together in the streamwise direction. Each 3D-PTV system consisted of one high-speed camera, ‘Mikrotron EoSens’, equipped with a four-way image splitter to mimic a classical four-camera setup, which allowed a continuous recording of 120 s.
\begin{table}
  \begin{center}
\def~{\hphantom{0}}
  \begin{tabular}{cccc}
      $ $  & $\boldsymbol{Ri0}$ & $\boldsymbol{Ri10}$ & $\boldsymbol{Ri20}$ \\[3pt]
       $U_{0}  [cm/s]$   & 10 & 10 & 10 \\[1ex]
       ${\Delta}\rho_{0}  [g/l]$   & 0 & 2.1 & 4.1 \\[1ex]
       $\alpha  [\degree]$  & 0 & 10 & 5 \\[1ex]
       $d [cm]$   & 5 & 5 & 5 \\[1ex]
       $Re_{0}=U_{0}d/\nu$ & 5000 & 5000 & 5000 \\[1ex]
       $Ri_{0}=g^{\prime}d\sin{\alpha}/U_{0}^{2}$ & 0 & 0.10 & 0.20 \\[1ex]
       $\eta [mm]$ & 0.23 & 0.29 & 0.31 \\[1ex]
       $L [cm]$ & 1.00 & 0.96 & 1.00 \\[1ex]
       $l_{sk} [cm]$ & 5.21 & 3.60 & 3.16 \\[1ex]
       $u_{\eta} [cm/s]$ & 0.43 & 0.34 & 0.32 \\ [1ex]
       $Re_{L} =u^{\prime} L/{\nu}$ & 152 & 107 & 102 \\ [1ex]
  \end{tabular}
  \caption{Overview of flow parameters for the three flow cases. The subscript \textit{0} indicates the inflow parameters.}
\label{table:tab1}
  \end{center}
\end{table}
The light source for illumination was a diode-pumped Nd-YLF laser (Quantronix, Darwin Duo 527 nm). As flow tracers, we used neutrally buoyant polyamide particles with a mean diameter of 100 $\mu m$ (manufactured by \textit{Evonik Industries}, Germany).\\
Each single 3D-PTV system covered an observation volume of about $9$ $cm$ x $9$ $cm$ x $4$ $cm$ in the $x$ (streamwise), the $y$ (wall-normal), and the $z$ (spanwise) directions, respectively. The fields of view of the individual PTV systems overlapped for about $2$ $cm$ to track particles continuously across the different observation volumes. The start of the measurement volume was located about $50$ $cm$ away from the inlet and covered roughly $31$ $cm$ in the streamwise direction (figure \ref{fig:fig2}). 
For each observation volume, it was possible to track up to 3000 particles simultaneously. This corresponds to a mean inter-particle distance of about $3.5$ $mm$, equivalent to roughly $10\eta$, with $\eta=(\nu^{3}/\epsilon)^{1/4}$ being the Kolmogorov microscale, where $\nu$ is the kinematic viscosity and $\epsilon=u^{\prime 3}/L$ is the local dissipation. Here, $u^{\prime}$ is the root mean square of the velocity fluctuation and $L$ is the integral length scale of the turbulence, evaluated as the integral of the autocorrelation function of the streamwise velocity along $x$. The turbulence level was quantified through the integral Reynolds number $Re_{L} =u^{\prime} L/{\nu}$ (Table \ref{table:tab1}). As can be observed, the stable stratification reduces the turbulence level. Reference length and time scales were evaluated at a height of about 7 $cm$ from the top wall, a location that is far from the wall but still sufficiently far from the strongly intermittent region close to the TNTI.
The spatial resolution achieved in our experiments is not sufficiently accurate to resolve the Kolmogorov scale. However, it was considered adequate for the purposes of the present work and a suitable compromise between a large enough spatial domain and spatial resolution. As shown by \citet{krug2017fractal}, the smallest convolutions of the TNTI are on the order of $10{\eta}$ and the TNTI geometry is therefore sufficiently captured by our measurements. The Lagrangian coherent structure extraction method explained below is based on vorticity. Given that the vorticity is somewhat under-resolved in our measurements, the extracted VLCSs represent those of a filtered velocity field, where we neglect the effect of Kolmogorov-sized eddies.
The time resolution was set to 250Hz, which over-samples about 20 times $\tau_{\eta}$, with $\tau_{\eta}=(\nu/\epsilon)^{1/2}$ being the Kolmogorov time scale.
We applied a temporal Savitzky-Golay filter with a span of 0.5$\tau_{\eta}$ to the velocity data. This reduced experimental noise due to position uncertainty of tracked tracer particles \citep{luthi2005lagrangian, wolf2012investigations}.\\   
A well-known feature of experimental particle tracking data is that particle trajectories have variable length and may be partly interrupted due to e.g. optical occlusions. However, the Lagrangian coherent structure extraction method explained below requires un-interrupted trajectories. We therefore interpolated the Lagrangian velocity data on an Eulerian grid with a spacing of $5{\eta}$. Subsequently we advected fluid particles numerically from these Eulerian velocity fields. A similar procedure has been applied for example by \citet{ouellette2012dynamical}. In figure \ref{fig:fig2}, we show samples of numerically computed fluid trajectories. To estimate the error of numerically calculated fluid particle trajectories, we used the longest measured trajectories and computed the root mean square (r.m.s.) distance between particle positions at the end of the trajectories. For one full flow through time of the entire volume, we obtained an acceptable value of about one Kolmogorov length scale.
\subsection{TNTI identification and local entrainment velocity}
Following previous work, the identification of the TNTI was done using a threshold on the enstrophy field, $\omega^{2}=\boldsymbol{\omega \cdot \omega}$ where $\boldsymbol{\omega}$ is the vorticity vector \citep[see e.g.][]{holzner2008lagrangian, krug2015turbulent}. The TNTI location is then defined by an iso-surface corresponding to a specific $\omega_{th}^2$ threshold. In the present investigation, the threshold was fixed at $\omega_{th}^2=2.5 s^{-2}$ just above the noise level of the data. This value is very close to those chosen by \citet{krug2015turbulent} for the same flow and by \citet{wolf2012investigations} for a turbulent jet. The local velocity propagation of the TNTI surface with respect to the fluid particles, the entrainment velocity $v_{n}$, was estimated using the direct approach, presented by \citet{wolf2012investigations}. In their approach, $v_{n}$ is computed from:
\begin{equation}
  v_{n}=v_{\omega_{th}^{2}}-v_{f},
\end{equation}
where $v_{\omega_{th}^{2}}$ is the local velocity of TNTI and $v_{f}$ is the local flow velocity. To determine $v_{\omega_{th}^{2}}$, we used the positions of $\omega_{th}^2$-isosurface at consecutive time steps. Similar to the velocity data, we also applied a temporal filter to the measured TNTI interface locations to remove occasional spurious outliers in the irrotational region.       
\subsection{VLCSs eduction}\label{subsection:extraction}
The detection of the Lagrangian coherent vortices is based on LAVD theory. We recall the definition of LAVD:
\begin{equation}
 LAVD_{t_{0}}^{t}(x_{0})=\int_{t_{0}}^t |\boldsymbol{\omega}(x(s;x_{0}),s)-\boldsymbol{\bar{\omega}}(s)| \mathrm{d} s.
\label{equation:lavd}
\end{equation}
where $\boldsymbol{\omega}$ is the vorticity along fluid trajectories, $\boldsymbol{\bar{\omega}}$ is its spatial average, and $x(t,x_{0})$ denotes the fluid trajectory starting at $x_{0}$ at time $t_{0}$. According to \citet{haller2016defining}, a rotational Lagrangian coherent vortex is defined as an evolving material domain filled with a nested family of tubular surfaces of $LAVD_{t_{0}}^{t}(x_{0})$ with outward decreasing LAVD values. The boundary of the VLCS is the outermost tubular level surface, whereas its center is the innermost member of the LAVD level-surface family. LAVD-based methods have been successfully applied in the past to 2D data of satellite oceanic velocity fields and  DNS of forced turbulence, as well as to 3D data of the ocean model ‘SOSE’ \citep{haller2016defining}. Prior applications of the detection method to three-dimensional data utilized the physics and geometry of the problem to simplify the extraction of the structures. For example, the 3D vortex extracted by \citet{haller2016defining} from the ‘SOSE’ model, is a single vertically-oriented structure.
In the present study, we implemented this method for multiple three-dimensional vortical structures extraction without a prior knowledge on the physics or geometry. Our algorithm can be described as a two-step procedure. In the first step, we compute one-dimensional curves representing the center of the structures and in a second step, we determine the boundaries of the VLCSs.\\ 
In \citet{haller2016defining}, the centers of VLCSs are defined by one-dimensional ridges of the LAVD field. In general, the computation of one-dimensional ridges in three dimensions is a challenging task. In the present work, we address this task by extending the 2D  ‘gradient climbing’ algorithm proposed by \citet{mathur2007uncovering} to three-dimensions. This algorithm uses the property that trajectories computed on the gradient field of a scalar quantity tend to accumulate along the ridges of the scalar field. The final position of these trajectories can be exploited to determine one-dimensional candidates for the ridges identification. Our ridges extraction algorithm is implemented as follows:\\
 
(i) For any initial time $t_{0}$, we determine narrow regions in the ridges neighborhood, where the magnitude of the ${\nabla}(LAVD)$ is higher than a pre-defined threshold and use points inside these regions as the initial conditions for computing numerically the solutions $\boldsymbol{x}_{0}(t)$ of the gradient dynamical system:\\
\begin{equation}
 \frac{d\boldsymbol{x}_{0}}{dt}={\nabla}LAVD_{t_{0}}^{t_{0}+T}(\boldsymbol{x}_{0}),
\label{equation:gradlavd}
\end{equation}
where $t$ denotes the time and ${\nabla}$ denotes the spatial gradient with respect to the initial position $\boldsymbol{x}_{0}$.
The solution $\boldsymbol{x}_{0}(t)$ takes the initial conditions to the closest ridge along the local gradient field of the LAVD.\\

(ii) For a given initial condition, the computation of the corresponding solution $\boldsymbol{x}_{0}(t)$ is stopped if the following two conditions hold: (a) the Hessian matrix ${\nabla}^{2}LAVD_{t_{0}}^{t_{0}+T}(\boldsymbol{x}_{0}(t))$ has at least two negative eigenvalues (a prerequisite for a point to be on a ridge), and (b) the angle between the eigenvector $\boldsymbol{e}_{t_{0}}^{t_{0}+T}(\boldsymbol{x}_{0}(t))$ corresponding to the smaller-in-norm eigenvalue of the Hessian matrix ${\nabla}^{2}LAVD_{t_{0}}^{t_{0}+T}(\boldsymbol{x}_{0}(t))$ and ${\nabla}LAVD_{t_{0}}^{t_{0}+T}(\boldsymbol{x}_{0}(t))$ shows no appreciable change (a sign of closeness to a nearby ridge). For large enough T, the eigenvector $\boldsymbol{e}_{t_{0}}^{t_{0}+T}(\boldsymbol{x}_{0}(t))$ will be approximately tangent to a ridge. \\

(iii) To determine the ridges candidates, we use the final positions of the solutions $\boldsymbol{x}_{0}(t)$ and select among them only points with a sufficiently close neighbor point. To this end, we use a pre-defined threshold on the distance between two points.\\

(iv) We then group together points belonging to the same ridge and order them. To order the points, we sort them in ascending order with respect to their $x$, $y$ and $z$ coordinates and select among the three sets, the one that minimizes the curve arc length.\\ 

(v) Finally, we smooth the ridges. By parametrizing their $x$, $y$ and $z$ coordinates with respect to the arc length, we fit them with a cubic smoothing spline.\\

\begin{figure}
  \centerline{\includegraphics[width=1\linewidth]{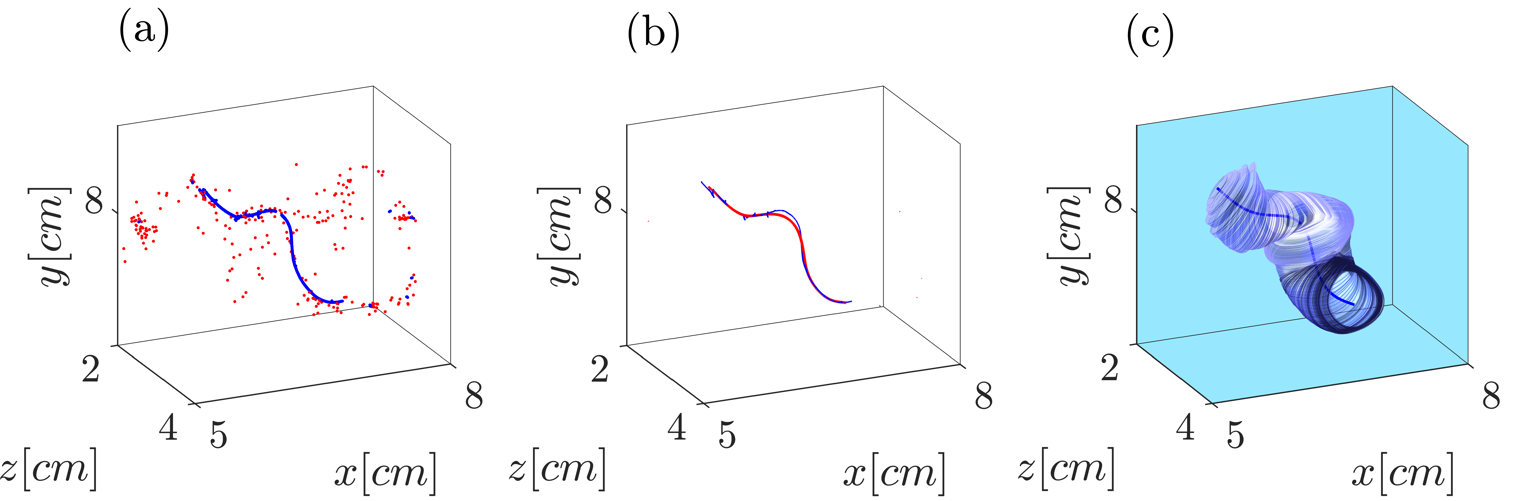}}
  \caption{Example of VLCSs extraction. In (a), the final position of the solution $\boldsymbol{x}_{0}(s)$ of equation \ref{equation:gradlavd} is shown. The blue dots are selected for the ridge construction, whereas the red ones are discarded. In (b), in blue, the connection of the points selected in (a) and its smooth fitting curve (red) are shown. The corresponding VLCS is shown in (c).}
\label{fig:fig3}
\end{figure}

In figure \ref{fig:fig3}, we show an example of the application of the last three steps described above. In this case, part of the points in ridge proximity are not aligned along a 1D curve (figure \ref{fig:fig3}(a)). We therefore select only points with sufficiently close neighbor points, sort them (blue curve in \ref{fig:fig3}(b)) and apply a smoothing cubic spline (long blue curve in \ref{fig:fig3}(c)).\\
After computing the structures centers, we determined their boundaries using the following steps:\\

(i) For each point of a given ridge, we erect point-wise normal planes to the ridge curve and determine the in-plane outermost almost-convex LAVD contour that encircles the point. These curves are one-dimensional curves in 3D.\\

(ii) We then use these curves to build the VLCSs boundaries. This is achieved for every pair of nearby curves by using the \textit{MATLAB} function $convhull$ to compute the lateral surface connecting them.\\

\begin{figure}
  \centerline{\includegraphics[width=1\linewidth]{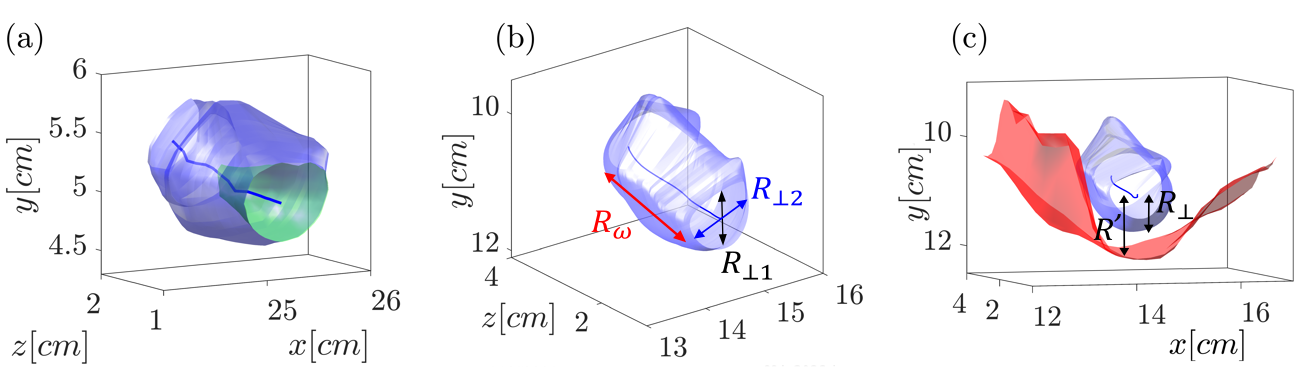}}
  \caption{(a) VLCS reconstruction following the algorithm of \citet{haller2016defining} (green inner-surface) and the modified algorithm introduced here (blue outer-surface). (b) Schematic of VLCS’s dimensions. (c) Schematic representation of  $R_{\perp}$, VLCSs (blue) cross-sectional size and $R^{\prime}$, the distance between VLCSs center and the TNTI  (red). }
\label{fig:fig4}
\end{figure}

In figure \ref{fig:fig3}(c), we show the result of the application of these steps to the ridges shown in figure \ref{fig:fig3}(b). The second step of the construction of the boundaries of the VLCS is slightly different from the one described in \citet{haller2016defining} in that these authors use tubular level surfaces of a fixed LAVD value. We observed that tubular LAVD level surface typically enclosed only part of the LAVD ridges. That is, although perfectly aligned to the structure’s center and enclosed by almost-convex contours, part of the ridge remained outside of the structure (see the example in figure \ref{fig:fig4}(a)), which is why we preferred to use the union of almost-convex contours. 
To give an impression on how the method performs on our data, we show several VLCSs in figure \ref{fig:fig2}(b). These are composed of tubular surfaces enclosing 1D-curves (centers), in the proximity of the TNTI, represented under the form of an open-surface. Here, the portion of the volume above the TNTI, where the VLCSs are located, corresponds to the turbulent part, whereas the lower side corresponds to the irrotational flow. While some of the structures lie ‘far’ from the TNTi, others are located close to it. The latter appear to shape the interface locally, as can be gleaned from the figure  \ref{fig:fig4}(c). This aspect will be investigated further in section \ref{subsec:interaction}.\\
In the following, we discuss the effect of the extraction time ${\Delta}t$ on the detected VLCSs, to explain how ${\Delta}t$ was chosen for the present data. We remark that for short extraction times, in the limit of ${\Delta}t \rightarrow 0$, VLCSs tend to their Eulerian counter parts \citep[see][]{haller2015lagrangian}. In this case, the material coherence is guaranteed only instantaneously, in the sense that there is no certainty that an Eulerian structure remains coherent over any observation time ${\Delta}t>0$. On the contrary, for very long extraction times, no-coherent structure can survive since the material coherence is limited in time for unsteady flows. For the vortical structures investigated here, the relevant temporal scale is the large eddy turnover time, which can be estimated as $t^{*}=L/u^{\prime}$.\\
For the measurement setup adopted in this work, there is a natural upper limitation for ${\Delta}t$. This is related to the residence time of a fluid particle in the observation volume (i.e. the time spent by a particle inside the measurement volume). For the portion of the measurement volume closer to the wall, we observed that the residence time varies between one and four turnover times. 
In order to set the extraction time in formula (\ref{equation:lavd}), we tested different ${\Delta}t$ values between zero (Eulerian proxy) and the maximum ($4t^{*}$) and investigated their effect in terms of $V_{VLCS}$, the average volume of a single VLCS. As can be observed in Table \ref{table:tab2}, $V_{VLCS}$ is weakly influenced by the extraction time, at least in the range  $t^{*}{\leq}{\Delta}t{\leq}4t^{*}$. The same applies to other properties related to the size, the shape and the orientation of the VLCSs (not shown). 
\begin{table}
  \begin{center}
\def~{\hphantom{0}}
  \begin{tabular}{cccccc}
      $ {\Delta}t/t^{*}$  & $ 0 $ & $ 1 $  &  $ 2 $ & $ 3 $ & $ 4 $ \\
       $V_{VLCS}/L^{3}$ & $ 0.28 $ & $ 0.41 $  &  $ 0.41 $ & $ 0.37 $ & $ 0.39 $ \\[1ex]
  \end{tabular}
  \caption{Average volume of the single VLCS, $V_{VLCS}$, for different extraction times of the flow case Ri20.}
\label{table:tab2}
  \end{center}
\end{table}
The extraction time has a considerable impact on the number of the structures that can be extracted using our measurement setup. As ${\Delta}t$ increases the number of trajectories entirely contained in the observation volume decreases drastically. This reduces the available volume for the VLCSs extraction and thus the number of the structures that can be educed. 
As a consequence of this and observing that the extraction time appears to influence only weakly the characteristics of VLCSs,  we opted to use ${\Delta}t=t^{*}$. We checked that qualitatively all results and conclusions remain the same for longer extraction times, albeit with reduced  statistics.
\subsection{VLCSs size and orientation}\label{subsection:sizeandori}
Given the three-dimensionality of VLCSs investigated here, we defined three characteristic dimensions (figure \ref{fig:fig4}(b)): one along the VLCS’s rotation axis and two in the cross-section. The two cross-sectional sizes are measured as follows. At each point of the center-line of the structure, we computed the point-wise perpendicular plane to the center-line. We then evaluated the intersection between this plane and the VLCS's boundary and we fitted an ellipse to the intersection points. By taking the average of the minor and major axes of the fitted ellipses, we assigned to each VLCS: $R_{{\perp}1}$ (the minor-) and $R_{{\perp}2}$ (the major cross-sectional size). The third dimension, $R_{\omega}$, is given by the length of the axis of rotation. As can be observed in figure \ref{fig:fig2}(b), some of the vortices are truncated in the rotation axis direction by the boundaries of the measurement domain. In such a case, we made an estimation of $R_{\omega}$ based on a quadratic fit. The fit was done in 1D using the average values between $R_{{\perp}1}$ and $R_{{\perp}2}$ along the rotation axis and using the zero crossing of the fitted curve. That is, we assumed that $R_{\omega}$  is finite and represented by the spatial distance between two cross-sections with zero area.\\  
The orientation of the VLCSs was assessed by computing the average unit vector $\textbf{n}$ tangent to the axis of rotation.\\
\section{Results}\label{sec:results}
\subsection{VLCSs geometrical properties}\label{subsec:VLCS gp}
\begin{figure}
  \centerline{\includegraphics[width=1\linewidth]{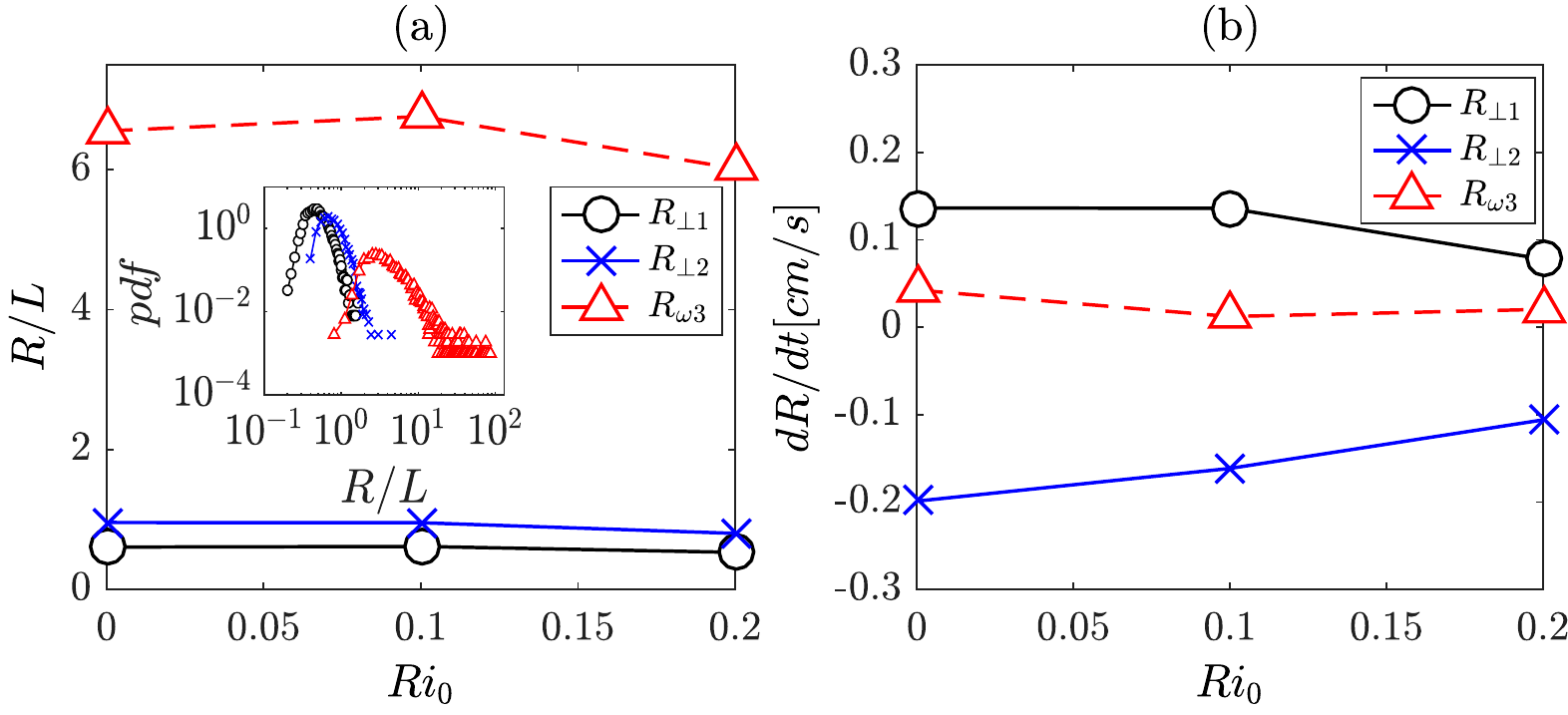}}
  \caption{Average VLCSs dimensions (a) and their growth rates (b) as a function of the initial Richardson number $Ri_{0}$.}
\label{fig:fig5}
\end{figure}

\begin{figure}
  \centerline{\includegraphics[width=1\linewidth]{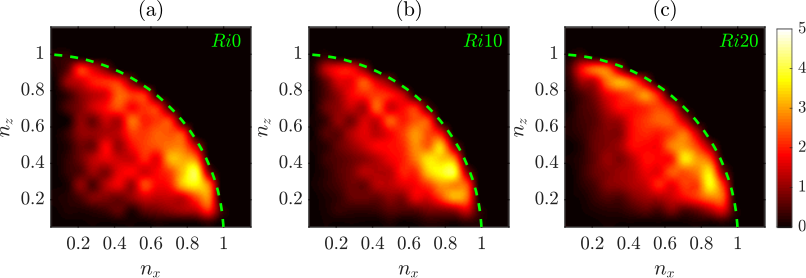}}
  \caption{Joint PDFs of VLCSs orientation in the $n_{x}$-$n_{z}$ plane at the initial time of the detection of the VLCSs for Ri0 (left), Ri10 (center) and Ri20 (right). The wall-normal component $n_{y}$ can be estimated from the joint PDFs reminding that $n_{x}^2+n_{y}^2+n_{z}^2\approx1$.}
\label{fig:fig6}
\end{figure}

\begin{figure}
  \centerline{\includegraphics[width=1\linewidth]{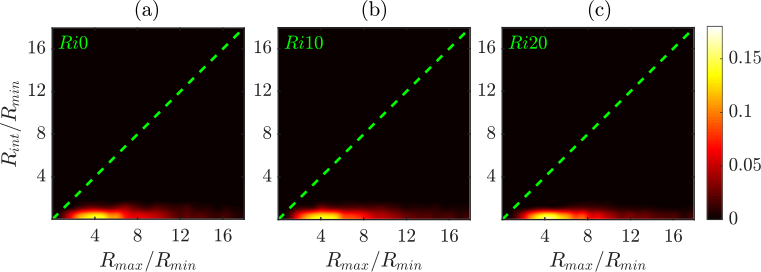}}
  \caption{Joint PDFs of VLCSs shape in the $R_{max}/R_{min}$ and $R_{int}/R_{min}$ map at the initial time of the detection of the VLCSs for Ri0 (left), Ri10 (center) and Ri20 (right).}
\label{fig:fig7}
\end{figure}

The average size parameters of the VLCSs as a function of the initial Richardson number $Ri_{0}$ are presented in figure \ref{fig:fig5}(a). Here, the three dimensions are normalized by the integral length scale of turbulence $L$, which is almost constant for all the flow cases (see Table \ref{table:tab1}). From figure \ref{fig:fig5}(a), together with the observation that $L$ is almost constant with $Ri_{0}$, it follows that the mean dimensions of the VLCSs do not vary significantly with the stratification. The cross-sectional average sizes $R_{{\perp}1}$ and $R_{{\perp}2}$ are equal to approximately $0.6L$ and $0.95L$. This gives an idea of the shape of the cross section of the structures, which on average is an ellipse with eccentricity of roughly $0.6$. The average size of the third dimension, the axis of rotation $R_{\omega}$, is of order $7L$. If the two cross-sectional sizes are ordered as a consequence of their construction, the third dimension is technically free to vary, i.e. it can be smaller or larger than $R_{{\perp}1}$ and $R_{{\perp}2}$. However, it is evident from the figure \ref{fig:fig5}(a) that the rotation axis of the VLCS is on average the longest one. The conclusion from this observation is that most of the structures appear to have a tubular shape. The inset in figure \ref{fig:fig5}(a), in which we show the probability density functions (PDFs) of the three size parameters for Ri20 gives an impression about their distribution.\\ 
The VLCSs dimensions presented in figure \ref{fig:fig5}(a) are measured at the initial time $t_{0}$ of their extraction. Solving the equation of motion between an initial time $t_{0}$ and a final time $t_{0}+{\delta}t$ for the particles constituting the centers and the boundaries of the VLCSs, it is possible to advect the structures under the flow evolution and to evaluate their final size, and hence their growth rates, $dR/dt$. Here, ${\delta}t$ was chosen to be equal to the extraction time. The growth rates of the cross-sectional sizes were evaluated computing $R_{{\perp}1}$ and $R_{{\perp}2}$ as described in the section \ref{subsection:sizeandori} at the final time $t_{0}+{\delta}t$. The growth rate of the axis of rotation was determined by the continuity equation, given that for a material structure $dR_{{\perp}1}/dt+dR_{{\perp}2}/dt+dR_{\omega}/dt=0$. An alternative way to determine $dR_{\omega}/dt$ is to compute it by directly estimating $R_{\omega}$ as described in section \ref{subsection:sizeandori} at $t_{0}$ respectively $t_{0}+{\delta}t$. However, we preferred the use of the continuity equation in order to avoid the inaccuracies introduced by the estimation approach of $R_{\omega}$ when the structure is not fully contained in the observation volume. In figure \ref{fig:fig5}(b), we display the average growth rates of the VLCS dimensions as a function of $Ri_{0}$. The growth rates corresponding to the minor-axis and to the rotation axis are positive in sign, and thus these axes increase their sizes in time, whereas the growth rate of the major axis is negative. The positive growth of the rotation axis implies a predominant stretching of the vortical structures along the rotation axis. In general, the picture that emerges is that under the flow evolution the VLCSs are typically stretched and their cross-section tends towards a more isotropic shape compared to their initial conditions.\\
Further, figure \ref{fig:fig5}(b) shows clearly that the growth rates diminish as $Ri_{0}$ increases. Thus, the stratification reduces the average VLCSs compression (of the intermediate axis) and stretching (of the smallest and the rotation axis dimensions). We also note that for all $Ri_{0}$ the magnitude of average growth rates is rather small and of the order of the Kolmogorov velocity scale magnitude $u_{\eta}$ (see Table \ref{table:tab1}), meaning that the VLCSs are on average not very strongly stretched. This result confirms our expectations, since by definition VLCSs are materially coherent structures that are not supposed undergo very significant deformation under flow evolution.\\  
In figure \ref{fig:fig6}, we present the orientation of the rotation axis of the VLCSs. Specifically, we plot the joint probability density functions (PDFs) of two components of the unit vector tangent to the axis of rotation of the VLCSs. Since there is no obvious choice in which direction the tangent vector should point, we show the absolute values of the two components.\\ 
For all the flow cases, the joint PDFs are biased towards values of $n_{x}\approx1$. That is, the structures exhibit a preferential orientation in the streamwise direction. Similarly, there is a sizeable probability to observe VLCSs oriented along the spanwise direction ($n_{z}\approx1$), whereas the probability of the wall-normal orientation ($n_{x}\approx n_{z}\approx0$) is not significant. As the Richardson number increases, the spanwise orientation gains some more importance at the expense of the streamwise one (figure \ref{fig:fig6}(c)).\\
To assess the shape of the structures, one can build a map of $R_{max}/R_{min}$ and $R_{int}/R_{min}$, with $R_{max}$,  $R_{int}$ and  $R_{min}$,  representing respectively the major, the intermediate and the minor VLCSs size. We should mention here that $R_{{\perp}1}$, $R_{{\perp}2}$ and $R_{\omega}$ do not coincide respectively with  $R_{min}$, $R_{int}$ and $R_{max}$ for all the structures, although we observed in figure \ref{fig:fig5} that this is true on average. The map is a useful tool to determine the shape of the VLCSs. In particular, values of $R_{max}/R_{min}$ and $R_{int}/R_{min}$ close to the origin $(1, 1)$ represent isotropic, sphere-like, structures. Values lying close to the abscissa denote tubular structures, whereas values in the proximity of bisector denote sheet-like structures. In figure \ref{fig:fig7}, we show joint PDFs of the shape map.
For all the flow cases, there is a clear prevalence of tubular structures, which persists with increasing stratification. The three joint PDFs show qualitatively similar behavior, with a peak of ($R_{max}/R_{min}$,$R_{int}/R_{min}$) between $(3, 1)$ and $(7, 1)$. The peak position is consistent with figure \ref{fig:fig5}(a), in which we showed that the average value of the rotation axis $R_{\omega}$ is about seven times larger than $R_{{\perp}1}$ and $R_{{\perp}2}$. 
\subsection{Interaction between the TNTI and VLCSs}\label{subsec:interaction}
\begin{figure}
  \centerline{\includegraphics[width=1\linewidth]{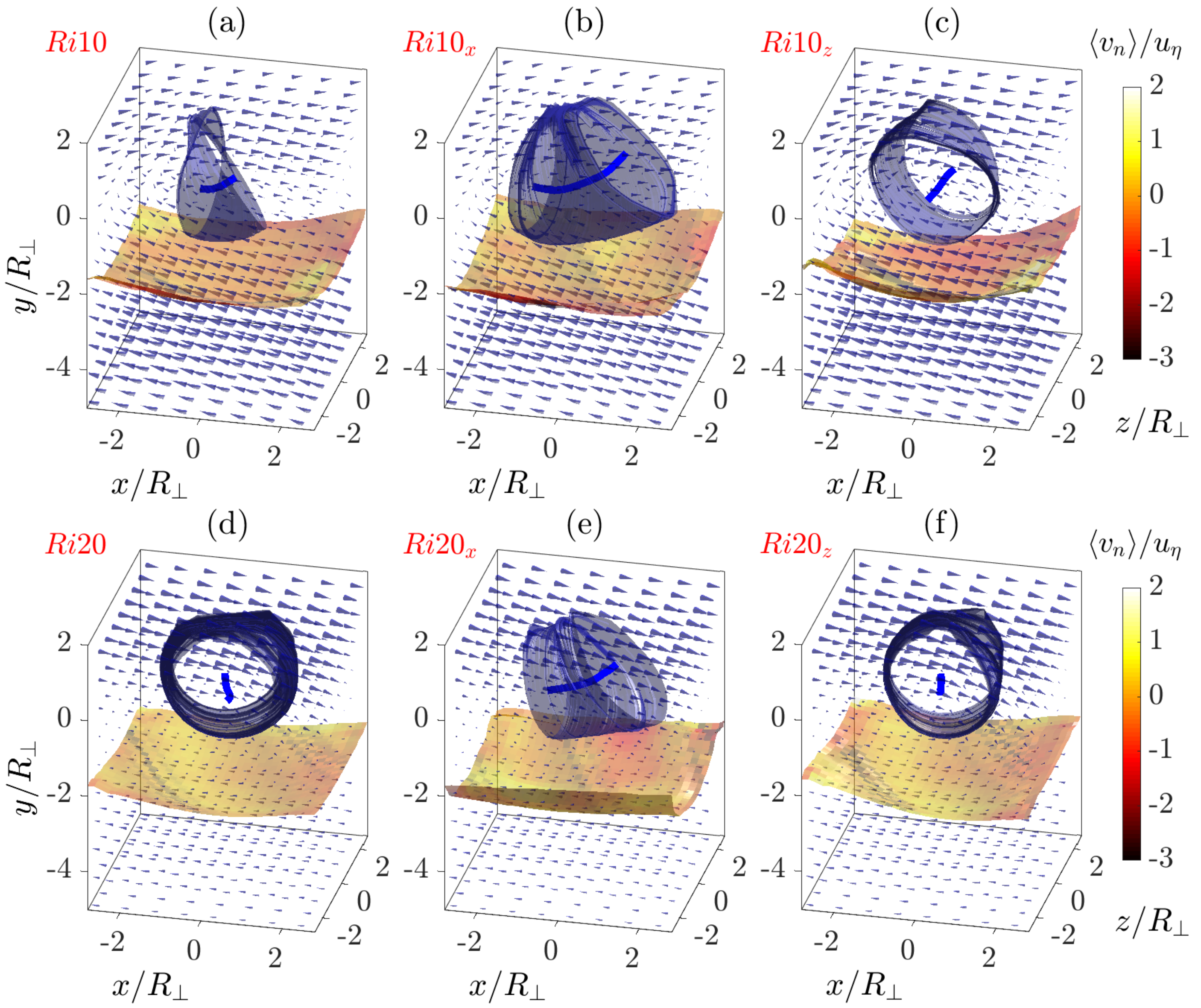}}
  \caption{Conditional average VLCS and TNTI position for Ri10 (upper-row) and Ri20 (lower-row). VLCSs centers are represented by the continuous blue lines and their boundaries by the tubular surfaces. The open surface is the conditioned TNTI mean position, color-coded with the average of the local entrainment velocity. The direction and the size of the vectors represent the conditional average velocity field. In the first column ((a) and (d)) the conditional analysis is made for all the structures, whereas in second ((b) and (e)) and the third column ((c) and (f)) the analysis is conditioned also on the orientation of the structures. The structures are oriented prevalently in the streamwise direction in (b) and (e), respectively in the spanwise direction in (c) and (f).}
\label{fig:fig8}
\end{figure}

In the following, we present the relationship between the TNTI and the nearby VLCSs. Through conditional analysis, we provide evidence that the average interface height and the local entrainment velocity are locally modulated by the presence of VLCSs. As observed in figure \ref{fig:fig2}(b), part of the VLCSs are located in the proximity of the TNTI. We selected VLCSs that are ‘sufficiently’ close to the TNTI by computing the ratio $r$ between $R^{\prime}$, the vertical distance of the center of the VLCS with respect to the TNTI, and $R_{\perp}$, the VLCSs cross-sectional average radius, defined as the one half of the mean value between $R_{{\perp}1}$ and $R_{{\perp}2}$. A sample representation of $R^{\prime}$ and $R_{\perp}$ can be found in figure \ref{fig:fig4}(c). Given that in the non-turbulent region there is no vorticity, the VLCSs cannot cross the TNTI. This implies that $r$ cannot be smaller than one. For the following conditional analysis, we selected structures with $r$ smaller than a threshold value $r_{th}=2.5$, which was fixed after testing different values and observing qualitatively similar results.\\
In a second step, for each selected structure we resampled both, the instantaneous velocity field at the initial extraction time and the LAVD field around it, onto a uniform grid. For this operation, we positioned the frame of reference at the center of the VLCS and normalized the three dimensions $x$, $y$ and $z$ around the structure with the VLCS’s cross-sectional average radius $R_{\perp}$. The rationale was to have a common frame of reference for all VLCSs and to compare flow fields, around VLCSs of the same normalized size. Taking the average of the LAVD fields around the VLCSs, we extracted a mean representative VLCS, that is, we applied the extraction algorithm described in \ref{subsection:extraction} to the average LAVD field.\\
Applying the same coordinate transformation to the TNTI surfaces in the proximity of the VLCSs, we computed the conditional average height of the TNTI. Moreover, at each location of the average height, we evaluated a mean local entrainment velocity ${\langle}v_{n}{\rangle}$. To this end, we computed the mean of instantaneous entrainment velocities near the structures. It is worth mentioning here that the high variance of the TNTI for the unstratified case Ri0 did not permit us to include this flow case in our analysis, given that the TNTI is observable in the measurement domain for a limited amount of instances, which did not allow us to obtain a meaningful statistical analysis.\\
In figure \ref{fig:fig8}, we present the results for Ri10 and Ri20. The centers of the structures are represented by the continuous lines close to the origin of the frame of reference and their boundaries by tubular surfaces enclosing them. Below the structures, the open surfaces represent the average TNTI positions, which we color-coded with the average local entrainment velocity. Around each structure, we show the direction of the average flow fields with cones that point along the velocity vector with the size representing its magnitude.\\
The first observation that emerges from the figure \ref{fig:fig8} is that the average VLCS is oriented differently for the two flow cases. For Ri10, the average VLCS is orientated in the streamwise direction (figure \ref{fig:fig8}(a)), whereas for Ri20, the VLCS is mainly oriented in the spanwise direction (figure \ref{fig:fig8}(d)).  
In both flow cases, the TNTI is positioned at about $y/R\approx-2$ and the surface is clearly modulated by the nearby structure, having a curvature that follows that of the VLCS’s boundaries. As the stratification increases, the curvature of the TNTI is observed to reduce, which is consistent with a decrease of the mean surface area of the TNTI.\\
In order to reveal the effect of the orientation of the VLCSs on the shape of the TNTI, we conditioned our analysis to streamwise (figure 8(b) and (e)), respectively spanwise oriented structures. To this end, we compared $n_{x}$ and $n_{z}$, evaluated as described in section \ref{subsection:sizeandori}. For a given VLCS, if $n_{x} >n_{z}$, the structure is considered to be oriented approximately in the streamwise direction, otherwise it is considered to be oriented in the spanwise direction. From  the second and the third column of the figure \ref{fig:fig8}, it appears clearly that the interface shape recalls that of the VLCSs boundaries having a larger curvature in the plane orthogonal to the rotation axis of the VLCS. Consider for example figure \ref{fig:fig8}(e), in which we conditioned our analysis to VLCSs of Ri20 oriented in the streamwise direction. The curvature of the average TNTI is almost entirely contained in $y-z$ planes, which are orthogonal to the center of the structure, while they are almost flat in the $x-y$ planes. Similarly, the curvature of the TNTI near the structures oriented prevalently in the spanwise direction is mostly limited to $x-z$ planes (see for example figure \ref{fig:fig8}(c)). 
The average entrainment velocity ${\langle}v_{n}{\rangle}$ is shown in color on the TNTI surface. As it is common practice, we normalized $v_{n}$ with the Kolmogorov velocity microscale $u_{\eta}$. Here, negative values of $v_{n}$ represent entrainment of dense irrotational fluid from below into the lighter turbulent fluid. The spatial distribution of ${\langle}v_{n}{\rangle}/u_{\eta}$ on the TNTI shows a similar pattern for the two flow cases in figure \ref{fig:fig8}, with higher negative values downstream with respect to the center of the structure, that is to say, close to $x/R_{\perp}\approx 2$ for the VLCSs oriented in the spanwise direction. For the structures oriented in the streamwise direction, ${\langle}v_{n}{\rangle}/u_{\eta}$ has higher negative values at the sides of the VLCSs near $z/R_{\perp}\approx\pm 2$. In correspondence of the center of the VLCS, for $(x/R_{\perp},y/R_{\perp})\approx(0,0)$  higher or even positive values of ${\langle}v_{n}{\rangle}/u_{\eta}$ are observed (see e.g. figure \ref{fig:fig8}(d)). The maximum negative value of ${\langle}v_{n}{\rangle}/u_{\eta}$ is different between the two flow conditions, diminishing (in terms of absolute value) for increasing stratification, from ${\langle}v_{n}{\rangle}/u_{\eta}\approx-1$ for Ri10 (figure  \ref{fig:fig8}(a)) to ${\langle}v_{n}{\rangle}/u_{\eta}\approx-0.5$ of Ri20 (figure \ref{fig:fig8}(b)). As previously observed, just below the center of the VLCSs, positive values of ${\langle}v_{n}{\rangle}/u_{\eta}$ can be noticed. The existence of regions of positive ${\langle}v_{n}{\rangle}/u_{\eta}$  (detrainment) is well known. \citet{wolf2012investigations} showed that $v_{n}/u_{\eta}$  can be positive in regions with positive curvature of the TNTI (concave curvature looking to the interface from the turbulent side). As seen in figure \ref{fig:fig2}, some of these bulges host VLCSs. As shown by others \citep[e.g.][]{watanabe2014enstrophy,krug2017global}, unconditioned averages of the ${\langle}v_{n}{\rangle}/u_{\eta}$ are negative (entrainment), but instantaneous positive (detrainment) values can be observed \citep{mistry2019kinematics}. To interpret the latter, one can take into account the local entrainment velocity decomposition based on the turbulent enstrophy equation introduced by \citet{holzner2011laminar}. Based on their decomposition, ${\langle}v_{n}{\rangle}/u_{\eta}$ can be locally positive if the enstrophy destruction outweighs both the enstrophy production, which is comparatively small in viscous superlayer, and the viscous diffusion, which is mostly positive in the viscous superlayer. This can lead to the reduction of the local enstrophy level below the threshold used for the TNTI identification.\\
In Table \ref{table:tab3}, we present the mean radius of curvature $R_{H}$ of the TNTI surfaces shown in figure \ref{fig:fig8}(a) and (d). 
\begin{table}
  \begin{center}
\def~{\hphantom{0}}
  \begin{tabular}{ccc}
      $ $  & $\boldsymbol{Ri10}$ & $\boldsymbol{Ri20}$ \\[3pt]
       $R_{H}/R_{\perp}$  & 6.6 & 9.4 \\[1ex]
       $\overline{{\langle}v_{n}{\rangle}}/u_{\eta}$  & -0.27 & -0.03 \\[1ex]

  \end{tabular}
  \caption{Average entrainment velocity and mean curvature of the surface of TNTI conditioned on the presence of VLCSs for Ri10(a) and Ri20(b).}
\label{table:tab3}
  \end{center}
\end{table}
\begin{figure}
  \centerline{\includegraphics[width=1\linewidth]{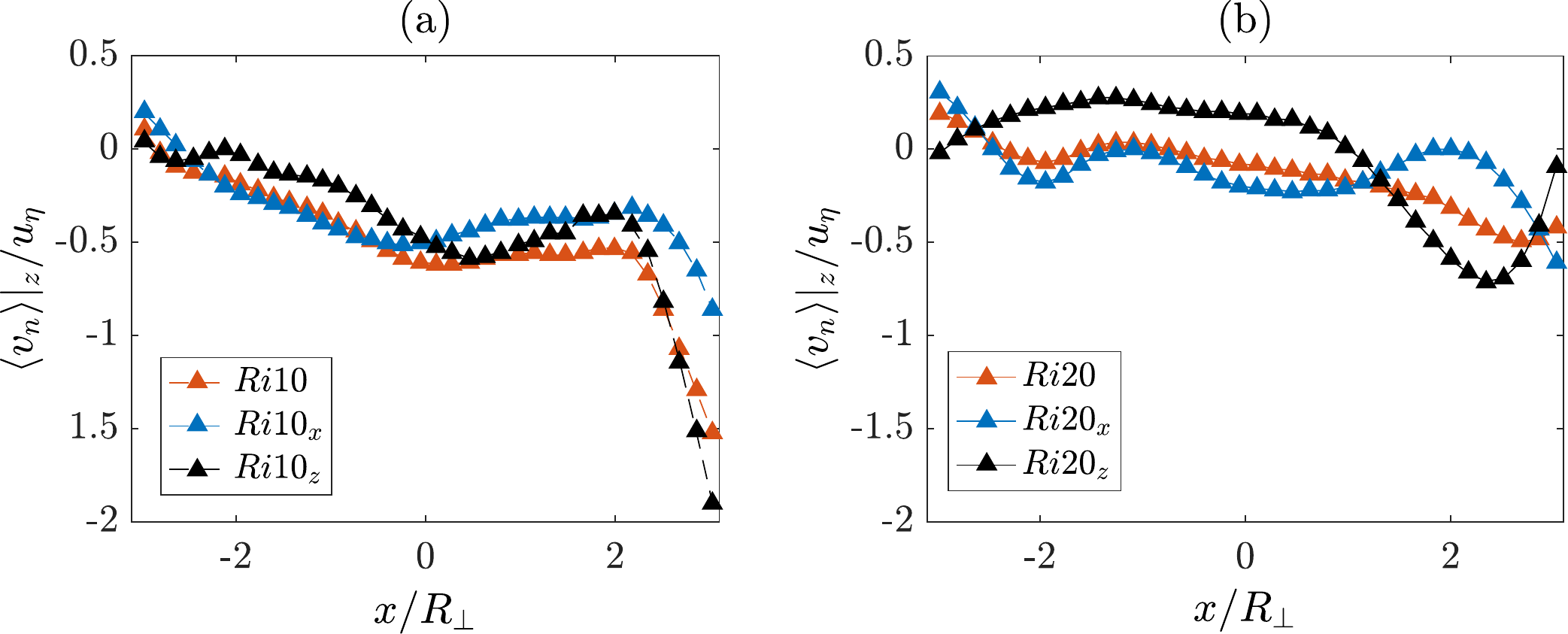}}
  \caption{Conditioned spanwise average of the local entrainment velocity in the proximity of the VLCSs related to the figures \ref{fig:fig8}(a) and (d).}
\label{fig:fig9}
\end{figure}
The mean radius of curvature increases from $R_{H}/R_{\perp}=6.6$ for Ri10 to $R_{H}/R_{\perp}=9.4$ for Ri20. The effectiveness of the VLCSs to contort the average interface reduces with increasing stratification. Although the mean radius of curvature is not a direct measure of the surface area of TNTI, it is clear that at higher values of $R_{H}$ correspond lower values of the surface area. It follows thus that the conditioned surface area of the TNTI decreases with increasing stratification, which is consistent with earlier work \citep[see e.g.][]{krug2015turbulent}. Furthermore, in Table \ref{table:tab3}, we report $\overline{{\langle}v_{n}{\rangle}}/u_{\eta}$ the average of the local entrainment velocity over the TNTI surfaces in figures \ref{fig:fig8}(a) and (d). The average of $\overline{{\langle}v_{n}{\rangle}}/u_{\eta}$ exhibit a higher value for the lower stratification passing from $\overline{{\langle}v_{n}{\rangle}}/u_{\eta}=-0.27$ for Ri10 to  $\overline{{\langle}v_{n}{\rangle}}/u_{\eta}=-0.03$ for Ri20.\\
\begin{figure}
  \centerline{\includegraphics[width=1\linewidth]{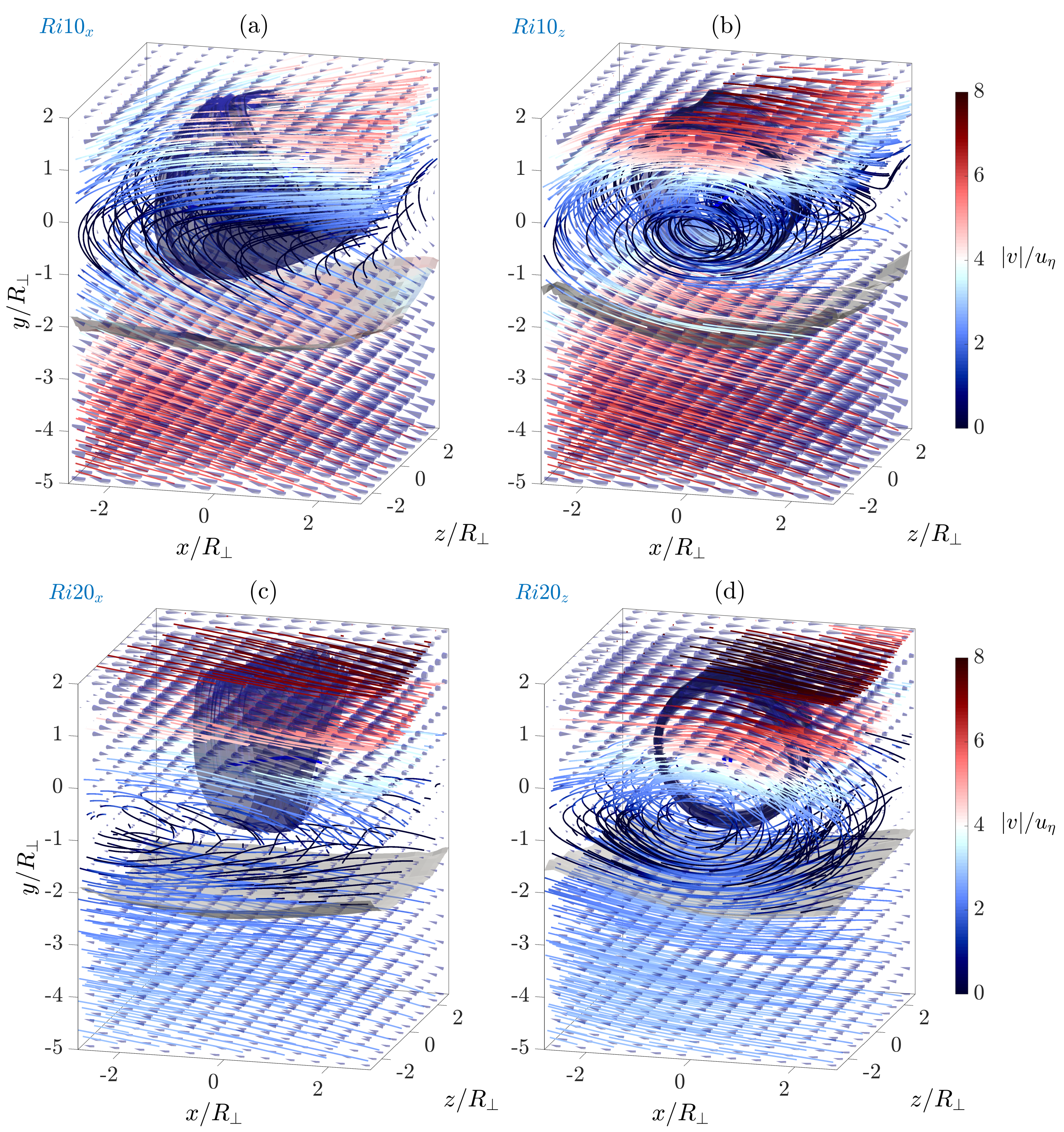}}
  \caption{Flow visualization. Streamlines of figure 8, color-coded with the average velocity magnitude.}
\label{fig:fig10}
\end{figure}
In order to further illustrate how the large-scale VLCSs influence the small scale entrainment, in figure \ref{fig:fig9} we show the the spanwise average of ${\langle}v_{n}{\rangle}/u_{\eta}$ corresponding figures \ref{fig:fig8}(a) and (d). In both cases shown in figure \ref{fig:fig9}, the entrainment velocity is higher in the downstream region ($x/R_{\perp}\approx 2$), and lower or even positive (figure \ref{fig:fig9}(b)) in the proximity of the center of the VLCS ($x/R_{\perp}\approx 0$). In a similar fashion to the figure \ref{fig:fig8}, we show the effect of the orientation of the structures on the entrainment velocity. For Ri10 (figure \ref{fig:fig9}(a)), it is clear that the entrainment has the same behavior for both, the structures oriented in the spanwise and in the streamwise directions. For Ri20 (figure \ref{fig:fig9}(b)), ${\langle}v_{n}{\rangle}|_{z}/u_{\eta}$ has considerably smaller negative values and for the structures oriented in the spanwise direction, it has positive values around 0.2 for $x/R_{\perp}\approx 0$.\\
Finally, we analyze how VLCSs near the TNTI influence the flow around them. The impact of the VLCSs on the mean flow in proximity of the TNTI surface results different for the two flow conditions shown in the figure \ref{fig:fig8}. For Ri10, no clear influence of the VLCs can be observed (figure \ref{fig:fig8}(a)). However, the spanwise oriented structures (figure \ref{fig:fig8}(c)) organize the flow both, inside and outside the turbulent zone. Inside the turbulent region, the average flow field results to revolve around the center of the structure, giving rise to a rotational motion, whereas outside, it is deviated towards the upstream region. In the case of Ri20, this behavior can be observed without the need of conditioning  on the orientation of the VLCSs (figure \ref{fig:fig8}(d)). However, this flow pattern is reinforced when only spanwise oriented structures are considered (figure \ref{fig:fig8}(f)).\\  
For a clearer visualization, we display in figure \ref{fig:fig10} the streamlines of the average flow fields around the conditionally oriented structures shown in figure \ref{fig:fig8}. Here, the streamlines are color-coded with the local velocity magnitude and the TNTI is represented by the gray-transparent open surface, positioned below the VLCS. For the structures oriented in the streamwise direction (figure \ref{fig:fig10}(a) and (c)), the streamlines in the non-turbulent zone appear to be rather horizontal, curving in the proximity of the VLCS ($y/R_{\perp}\approx 0$) and following the direction of the mean flow in the turbulent region. In both cases, the magnitude of the velocity field is higher far from the TNTI, both on the non-turbulent and the turbulent regions. When the spanwise oriented structures are considered (figure \ref{fig:fig10}(b) and (d)), a different flow organization arises. Outside the turbulent zone, far from the TNTI, the streamlines are again almost horizontal, similar to those close to the streamwise oriented structures in figure \ref{fig:fig10}(a) and (c). However, in the turbulent side, they follow the rotational motion induced by the VLCSs, curling up around the structures. This is evident in the figures \ref{fig:fig10}(c) and (d), where the swirling motion due to the presence of the structures can be clearly distinguished. For Ri20 (figure \ref{fig:fig10}(d)), the streamlines follow almost tangentially the TNTI. The velocities along the streamlines forming the swirling motion inside the turbulent zone are higher for both Ri10 and Ri20 on the upper side of the structures ($y/R_{\perp}\approx\pm 1$), decreasing in the proximity of the center of the structures and increasing again in the non-turbulent side.  
\section{Discussion and summary}\label{sec:discussion}
In this paper, we focused on the detection and characterization of Lagrangian vortical coherent structures (VLCSs) and their influence on the turbulent/non-turbulent interface (TNTI) and entrainment of a gravity current. Using 3D-PTV data, the VLCSs were educed by means of the so-called Lagrangian-Averaged Vorticity Deviation (LAVD) method. The TNTI was identified using an enstrophy threshold, whereas its entrainment velocity was computed through a direct method described in \citet{wolf2012investigations}.\\
In the section \ref{subsec:VLCS gp}, we described the geometrical characteristics of the VLCSs. In particular, in figure \ref{fig:fig5}(a) we observed that the average cross-sectional dimensions of the VLCSs are of order of the integral length scale of the turbulence $L$. By normalizing them with $L$, almost no-variation of their size with increasing stratification was noticed. Thus, the size of the VLCSs appeared to scale with integral length scale. A similar observation was made for the largest vortical structures near the TNTI of a turbulent jet by \citet{da2011role}. Using a low-pressure iso-surface for the structures eduction, the authors found that the radius of what they call large-scale vortical structures is of order of Taylor microscale. Furthermore, analyzing the growth rates of the dimensions of the VLCSs, we noticed that VLCSs are predominantly stretched and in time, their cross-sections tend towards a rather isotropic shape. This is reminiscent of the predominant vortex stretching mechanism \citep{tsinober2000vortex}, which is well known e.g. since the initial studies by \citet{chong1990general}, \citet{cantwell1993behavior} and \citet{soria1997volume} on the invariants of velocity gradient tensor. Through coarse grained and filtered velocity gradient tensors, \citet{meneveau2011lagrangian} demonstrated that predominant stretching is discernible also at larger flow scales that are well in the inertial range, as is it the case for the ones investigated here.\\ 
In figure \ref{fig:fig6} and \ref{fig:fig7}, we showed that on average the VLCSs are of tubular-shape orientated mainly in the streamwise direction. The fact that the structures are prevalently oriented in the streamwise direction is interesting, given that in our flow, the mean vorticity is oriented in the spanwise direction. A well known picture in wall-bounded turbulence is that an initially spanwise oriented vortex, formed near the wall of the boundary layer, is disturbed by an ejection event that rises part of the vortex tube at height where the mean flow is faster. The mean flow advects this coherent mass faster than the vortex tube near the wall tilting its legs towards vertical planes, in which they are stretched by the mean shear \citep{kim1999very}. We speculate that a similar mechanism may be at the base of formation of the VLCSs investigate here. In the mixing layer of the gravity current, initial vortices form via a Kelvin-Helmholtz type mechanism and are then tilted by turbulence and the mean shear. In figure \ref{fig:fig6}, we also noted that as the stratification increases, more structures tend to be oriented in spanwise direction. We associate this to the mechanism described before. Indeed, as the stratification increases the vertical motion of the fluid is known to be reduced. This attenuates sweeps and ejections, with the consequence that the probability to observe spanwise oriented structures may be higher. Moreover, the orientation of the structures close to the TNTI was shown to be almost horizontal. This is a consequence of the fact that the VLCSs cannot cross the interface and cannot finish or start on it. This is in line with the findings of \citet{da2011role} in the case of a planar turbulent jet.\\
In subsection \ref{subsec:interaction}, we investigated the interaction between large scale VLCSs and the TNTI, with a focus on both elements constituting the entrainment process, namely, the TNTI area and the local entrainment velocity. We showed that the VLCSs modulate the TNTI height, thereby increasing the TNTI surface area. A similar observation was done by \citet{lee2017signature} for the TNTI height of a TBL. Here the authors conducted a conditional analysis based on LSMs position, showing that the interface is locally contorted by the LSMs. In both examples, the gravity current and the TNL, it is demonstrated that the large-scale flow structures enhance the TNTI area thereby augmenting the entrainment flux. Moreover, we showed that the local entrainment velocity at the smaller scales of the turbulence is modulated by the large scale VLCSs (figure \ref{fig:fig8}). In particular, the local entrainment velocity was seen to be higher downstream with respect to the position of the VLCSs, decreasing and becoming even positive (detrainment) just beneath the center of the structure. We hypothesize that this might be connected to presence of the VLCS, which induces a motion tangent to the surface of the TNTI locally reducing the entrainment rate. The visualization of streamlines of the mean velocity field supports this idea. A similar remark was made by \citet{bisset2002turbulent} for the instantaneous streamlines near a bulge of the TNTI of a turbulent wake. Here the authors observed that the streamlines enter it the turbulent side (high entrainment) only in regions with a convex curvature of the surface as seen from the turbulent side \citep[see figure 15 in][]{bisset2002turbulent}, whereas beneath the bulge the streamlines are almost horizontal (low entrainment or detrainment). In figure \ref{fig:fig2}, we observed that part of these bulges hosts a VLCS, which is compatible with findings in \citet{bisset2002turbulent}. A more recent work by \citet{mistry2019kinematics}, which discusses the existence of instantaneous detrainment zones in a turbulent jet, further supports our observations on the detrainment near the VLCSs. Here, the authors show that similarly to our findings high detrainment is observed when the fluid moves tangentially to the interface on both the sides of the TNTI.\\

We are grateful for financial support from DFG priority program SPP 1881 under grant number HA 7497/1-1.

\bibliographystyle{jfm}

\end{document}